\documentclass[aps, prb, superscriptaddress, twocolumn, longbibliography, showpacs, preprintnumbers,
amssymb, floatfix]{revtex4-1}

\usepackage{graphicx}
\usepackage{epsfig}
\usepackage{amsmath}
\usepackage{amssymb}
\usepackage{amsbsy}
\usepackage{xcolor}
\usepackage{soul}
\usepackage{hyperref}

\newcommand{\spvec}[1]{\ensuremath{\mathbf{#1}}}
\newcommand{\unitvec}[1]{\ensuremath{\mathbf{\hat{#1}}}}

\newcommand{\colvec}[1]{\ensuremath{\mathrm{#1}}}

\newcommand{\commentout}[1]{{}}

\newcommand{\beq}{\begin{equation}}
\newcommand{\eeq}{\end{equation}}

\newcommand{\cbE}{\boldsymbol{\mathbf{\cal E}}}
\newcommand{\cbH}{\boldsymbol{\mathbf{\cal H}}}

\begin{document}

\title{Cooperative field localization and excitation eigenmodes in disordered metamaterials}
  \date{\today }

\author{Nikitas Papasimakis}
\affiliation{Optoelectronics Research Centre and Centre for Photonic Metamaterials, University of Southampton, Southampton SO17 1BJ, United Kingdom}
\author{Stewart D. Jenkins}
\affiliation{Mathematical Sciences and Centre for Photonic Metamaterials, University of Southampton, Southampton SO17 1BJ, United Kingdom}
\author{Salvatore Savo}
\affiliation{Optoelectronics Research Centre and Centre for Photonic Metamaterials, University of Southampton, Southampton SO17 1BJ, United Kingdom}
\affiliation{TetraScience Inc., 114 Western Ave, Boston, MA, 02134}
\author{Nikolay I. Zheludev}
\affiliation{Optoelectronics Research Centre and Centre for Photonic Metamaterials, University of Southampton, Southampton SO17 1BJ, United Kingdom}
\affiliation{Centre for Disruptive Photonic Technologies, School of Physical and Mathematical Sciences and The Photonics Institute,
            Nanyang Technological University, Singapore 637378, Singapore}
\author{Janne Ruostekoski}
\affiliation{Mathematical Sciences and Centre for Photonic Metamaterials, University of Southampton, Southampton SO17 1BJ, United Kingdom}
\affiliation{Department of Physics, Lancaster University, Lancaster, LA1 4YB, United Kingdom}

\begin{abstract}
We investigate numerically and experimentally the near-field response of disordered arrays comprising asymmetrically split ring resonators that exhibit strong cooperative response.
Our simulations treat the unit cell split ring resonators as discrete pointlike oscillators with associated electric and magnetic point dipole radiation, while the strong cooperative radiative coupling between the different split rings is fully included at all orders.
The methods allow to calculate local field and Purcell factor enhancement arising from the collective electric and magnetic excitations. We find substantially increased standard deviation of the Purcell-enhancement with disorder, making it increasingly likely to find collective excitation eigenmodes with very high Purcell factors that are also stronger for magnetic than electric excitations. We show that disorder can dramatically modify the cooperative response of the metamaterial even in the presence of strong dissipation losses as is the case for plasmonic systems. Our analysis in terms of collective eigenmodes paves a way for controlled engineering of electromagnetic device functionalities based on strongly interacting metamaterial arrays.
\end{abstract}
\maketitle


\section{Introduction}
The propagation of waves through disordered media is a ubiquitous theme across diverse research areas, from electrodynamics and solid state physics to acoustics and fluid mechanics. In the field of optics, in particular, the quest for utilizing disorder and analyzing its effects is attracting considerable interest from the fundamental studies of transport phenomena to potential applications, such as random lasing~\cite{WiersmaNatPhys2008}, hyper-transport~\cite{hypertrans}, and image transport through optical fibers~\cite{imagetrans}. While metamaterials have so far been almost solely based on periodically structured resonator arrays, there is an increasing interest in extending these also to the realm of  disordered systems, where disorder is introduced either in the form of inhomogeneous broadening~\cite{PhysRevE.73.056605,Zharov2005,Gollub,JenkinsRuostekoskiPRB2012b,Ustinov2015,Fistul2017} or as random perturbations in the resonator positions~\cite{Aydin2004,papasimakis2009,Helgert2009,Singh2010,Engheta2010,Albooyeh2010,SavoEtAlPRB2012,Zhang2017,Marini2016,Albooyeh2014,Jang2018,Pinheiro2017}. Whereas the former can affect the strength of interactions between the resonators, the latter can lead to qualitative changes in the response of the resonator array~\cite{JenkinsRuostekoskiPRB2012b}. Recent work has also included suggestions for a number of applications, such as  topological photonics \cite{Zhang2017}, random lasing with gain~\cite{Marini2016}, perfect absorbers \cite{Giessen2014,Giessen2016} and wavefront shaping \cite{Albooyeh2014,Arango2015,Jang2018}.

In this work, we study the near-field response of positionally disordered metamaterial arrays consisting of asymmetrically split rings (ASRs)~\cite{FedotovEtAlPRL2007,papasimakis2009,SavoEtAlPRB2012}.
Our approach includes numerical simulations of the full metamaterial array, where each meta-atom is considered individually, and experimental near-field measurements. The interplay between disorder and the strong inter-metamolecule interactions results in radical changes to the metamaterial response as compared to that of a regular array. Indeed, examining the collective radiative excitation eigenmodes of regular and disordered metamaterials reveals striking changes in their EM response.
 A regular planar array of ASR metamolecules can support a giant, spatially extended subradiant excitation, where most of the excitation occupies a single collective eigenmode~\cite{Jenkins2017PRL}.
 Even small amounts of disorder, weakly perturbing the metamolecule positions, can strongly localize the eigenmodes, and this change is directly reflected in the  far-field response~\cite{Jenkins2018}.
 Here we show that in the near field, the localized excitation energies of both electric and magnetic dipoles grow with increasing disorder, eventually saturating, and in the case of magnetic dipoles, finally decreasing at large values of disorder. The field confinement due to disorder is described in terms of the Purcell factors.  We find that, in particular, the standard deviations of the maxima of the Purcell factors over collective modes and stochastic realizations substantially increase with disorder. For strong disorder it is increasingly likely that there are collective eigenmodes with very high Purcell factors. Our findings indicate that this sensitivity of the cooperative response on disorder strength depends heavily on dissipation losses. In the case of low-loss (microwave) ASR arrays, manifestations of disorder-induced collective phenomena are readily observed. On the other hand, careful engineering of the metamolecule properties allows the observation of such phenomena even in the case of plasmonic, lossy metamaterial systems operating in the optical part of the spectrum.

Our analysis is focussed on the fundamental understanding of the cooperative microscopic principles of the macroscopic EM response and paves a way to overcome the deleterious and unwanted effects of disorder in order to benefit from them. Modifying collective interaction phenomena in a controlled way provides a platform for harnessing and engineering complex disorder-dependent EM-field response for the design of metamaterial-based devices with prescribed functionalities. By means of decomposing the excitations into eigenmodes that can be shaped and designed by adjusting the disorder, we demonstrate that one can engineer the near-field landscape and, e.g., selectively prepare desired localized multipole (such as magnetic dipole) excitations. In particular, the number of eigenmodes required to achieve the target states is decreasing with increasing disorder. Hence, one needs to employ only a handful of modes to engineer localized excitations in strongly disordered arrays.

\section{Collective response and numerical model}
\label{section:model}
We utilize the theoretical model based on coupled dipolar scattering centers that we have developed first for regular arrays~\cite{JenkinsLongPRB,JenkinsLineWidthNJP,CAIT} and recently generalized for disordered cases \cite{Jenkins2018}. The approach is suitable for simulations of a cooperative response~\cite{Lehmberg1970a,Lehmberg1970b,Ishimaru1978,Ruostekoski1997a,JavanainenMFT} in large, strongly coupled, magnetodielectric resonator arrays, while closely related models based on point-dipole scatterers can be used, e.g., in atomic arrays~\cite{Jenkins2012a,Facchinetti}. Other point-dipole scatterer techniques that utilize similar principles have more recently been applied in the design and modelling of metasurfaces \cite{Drsmith2017,Drsmith2017b}.

The studies both in regular and disordered arrays provide a good qualitative agreement with the experiments, indicating that the essential features of the collective responses in these systems can be captured by accurate descriptions of the field-mediated interactions between the scattering centers even when the microscopic features of the resonators are only approximately incorporated in the point dipole model.

\begin{figure}
  \centering
  \includegraphics[width=1\columnwidth]{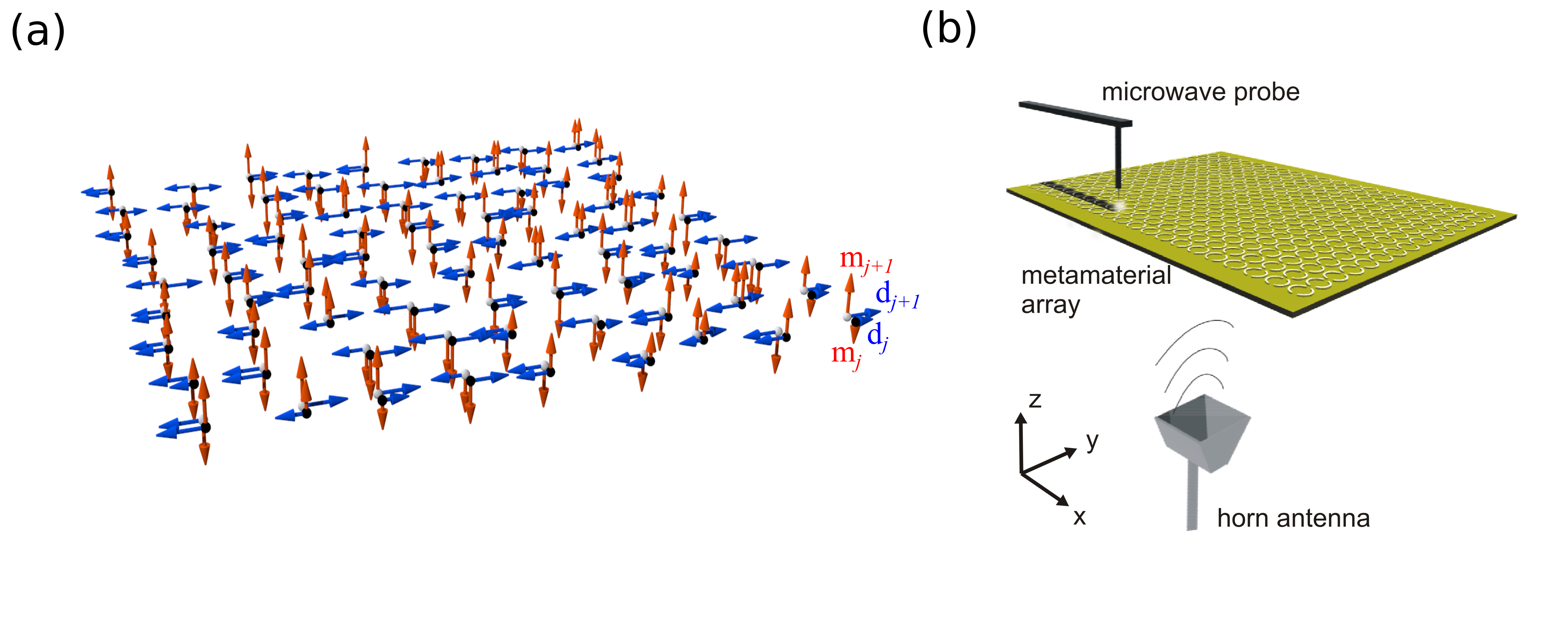}
  \caption{
    \textbf{a. Theoretical model for positionally disordered asymetrically split ring (ASR) arrays.} Each meta-molecule consists of two arc resonators (meta-atoms) which are represented by grey and black spheres. Blue and red arrows show the electric ($d_\ell$) and magnetic ($m_\ell$) dipole moment of each meta-atom under plane wave illumination.
    \textbf{b. Experimental setup for near-field characterisation of microwave ASR metamaterial.}
    The resonators are arranged in a square lattice with lattice spacing
    $a = 7.5$ mm.  The inner and outer radii of each ASR are $2.8$ and
    $3.2$ mm respectively. The ASR array is supported by a FR4 dielectric substrate. A broadband linearly polarized horn antenna illuminates the
    sample, and a microwave monopole antenna measures the electric field near the surface
    of the array.
  }
  \label{fig:schematic}
\end{figure}

We briefly highlight the main elements of the theory~\cite{JenkinsLongPRB}. Each ASR meta-molecule in a 2D array is labelled by index $\ell = 1 \ldots 30\times 36$. The dominant effect of the excitations is described by
the amplitudes $d_\ell$ and $m_\ell$, where the symmetric oscillations possess a net electric dipole proportional to $d_\ell \unitvec{d}$ and anti-symmetric current oscillations a net magnetic dipole $m_\ell \unitvec{m}$ with a small electric quadrupole \cite{FedotovEtAlPRL2007} (see Fig.~\ref{fig:schematic}a).
In order to model the effect of spatial disorder, the ASR $\ell$ is assumed to be located at position $\spvec{r}_\ell = \spvec{R}_\ell + \delta\spvec{r}_\ell$, where $\spvec{R}_\ell$ is the center of the
corresponding unit cell, and $\delta\spvec{r}_\ell$ is the random displacement of the ASR.
Each unit-cell resonator is decomposed into two asymmetric arcs, or meta-atoms, the excitations of the arcs are described by the oscillator normal mode amplitudes $b_j$.
For simplicity, for the unit-cell excitations we use a normalization of $d_\ell$ and $m_\ell$ for which the
lower arc of unit cell $\ell$ has the amplitude $b_{2\ell-1} = (d_\ell +i m_\ell)/\sqrt{2}$, and the
amplitude of the upper arc $b_{2\ell} = (d_\ell - im_\ell)/\sqrt{2}$. The total energy contained in an ASR excitation is proportional to $|d_\ell|^2 +
|m_\ell|^2$.
Throughout the discussion we assume that all field and resonator amplitudes refer to the slowly-varying versions of the
positive frequency components of the corresponding variables,
where the rapid oscillations $e^{-i\Omega t}$ ($k=\Omega/c$) due to the frequency, $\Omega$, of the incident wave have been factored out in the rotating wave approximation.

In the numerical implementation, each arc (meta-atom) $j$ ($j=1\ldots 2N$) behaves like a damped $RLC$ circuit driven by external fields and the fields scattered by the other arcs.
Oscillations in every arc are damped at rate $\Gamma=\Gamma_e+\Gamma_m+\Gamma_o$, where the electric and magnetic dipole radiation, and non-radiative Ohmic loss rates are $\Gamma_e$,
$\Gamma_m$ and $\Gamma_o$, respectively. In the case of microwave ASR resonators ohmic loss occurs mainly in the dielectric substrate and is represented by $\Gamma_o=0.07\Gamma$, while in the case of plasmonics, losses occur mainly in the metal and are set at $\Gamma_o = 0.25 \Gamma$~\cite{Jenkins2018}. To enhance the strength of cooperative interactions, we consider realistic
arrays of metallic meta-molecules that are closely spaced with a lattice spacing of $a=0.28\lambda$ and $a=0.2\lambda$ for microwave and plasmonic metamaterials, respectively.
With the symmetry of the problem we obtain $\spvec{d}_j(t) = d_j(t)\unitvec{e}_y$  and
magnetic dipole $\spvec{m}_j(t) = m_j(t)\unitvec{m}_j$, where $\unitvec{m}_{2\ell} = -\unitvec{m}_{2\ell-1} \equiv \unitvec{m} = \unitvec{e}_z$.
The upper and lower arcs are located at $\spvec{r}_j +
(u/2)\unitvec{e}_y$ and $\spvec{r}_j - (u/2)\unitvec{e}_y$,
respectively, where $u$ denotes the parameter representing the size of the unit cell. If the split rings were symmetric, the individual meta-atoms would have identical resonance frequencies $\omega_j = \omega_0$, while an asymmetry in the arc lengths shifts the meta-atom resonance frequencies by $\delta\omega$ so that for ASR $\ell$
$ \omega_{2\ell-1} = \omega_0 - \delta\omega$ and
 $\omega_{2\ell} = \omega_0 + \delta\omega $.

The dynamics of the meta-atom $j$ follows from the fact that it is driven by the incident fields, $\cbE_0(\spvec{r},t)$ and $\cbH_0(\spvec{r},t)$,  and the sum of the fields ${\bf E}_{S}^{(l)}({\bf r},t)$ and ${\bf H}_{S}^{(l)}({\bf r},t)$ scattered by all the other resonators $l$ in the system,
\begin{align}
{\bf E}_{\text{ext}}({\bf r}_j,t)& = \cbE_0(\spvec{r},t)
+ \sum_{l\ne j}{\bf E}_{S}^{(l)}({\bf r},t),
\label{eq:Eext}\\
{\bf H}_{\text{ext}}({\bf r}_j,t)& = \cbH_0(\spvec{r},t) +
\sum_{l\ne j}{\bf H}_{S}^{(l)}({\bf r},t)\,,
\label{eq:Bext}
\end{align}
where the scattered field contributions  from the meta-atom $l$ read
\begin{align}
{\bf E}_{S}^{(l)}({\bf r},t)&=\frac{k^3}{4\pi\epsilon_0}
\bigg[{\sf G}({\bf r} -{\bf r}_l){\bf d}_l
+\frac{1}{c}{\sf G}_\times({\bf r} - {\bf r}_l)
{\bf m}_l\bigg],\label{eq:Esc}\\
{\bf H}_{S}^{(l)}({\bf r},t)&=\frac{k^3}{4\pi}
\bigg[{\sf G}({\bf r} -{\bf r}_l){\bf m}_l
-c{\sf G}_\times({\bf r} - {\bf r}_l){\bf d}_l\bigg]\,.
\label{eq:Bsc}
\end{align}
The dipole radiation of
the electric (magnetic) field at ${\bf r}$, from an oscillating electric (magnetic) dipole with an amplitude $\hat{\bf d}$
at the origin is given by~\cite{Jackson}
\begin{align}
{\sf G}&({\bf r})\,\hat{\bf d}=
(\hat{\bf n}\!\times\!\hat{\bf d}
)\!\times\!\hat{\bf n}{e^{ikr}\over kr}\nonumber\\
& +[3\hat{\bf n}(\hat{\bf n}\cdot\hat{\bf d})-\hat{\bf d}]
\bigl[ {1\over (kr)^3} - {i\over (kr)^2}\bigr]e^{ikr}
-{4\pi \hat{\bf d}\,\delta(k {\bf r})\over3}\,,
\label{eq:DOL}
\end{align}
where $\hat{\bf n} = {{\bf r}/ r}$. Similarly, the electric (magnetic) field at ${\bf r}$ of an oscillating magnetic (electric) dipole with an amplitude $\hat{\bf d}$
at the origin is
\beq
{\sf G}_\times ({\bf r})\,\hat{\bf d}= {i\over k} \nabla\times {e^{ikr}\over kr}\, \hat{\bf d}\,.
\eeq

The radiative field-mediated interactions lead to a coupled set of linear equations, describing the dynamics of  the normal mode amplitudes of the arc variables $\colvec{b} \equiv (b_1, b_2, \ldots, b_{2N})^T$~\cite{JenkinsLongPRB}, where
(unnormalized) $b_j$ of each meta-atom is given in terms of its electric and magnetic dipoles
\beq
 b_j(t) =
 \sqrt{\frac{k^3}{12\pi\epsilon_0}}
 \Bigg[
\frac{d_j}{\sqrt{\Gamma_{e}}}
  +
 i \frac{m_{j}}{c\sqrt{\Gamma_{m}}}
 \Bigg]
 \,\text{.}
 \label{eq:b_multipole}
 \eeq
The system of $N$ ASR meta-molecules ($2N$ single-mode resonator arcs) possesses $2N$ collective eigenmodes of current oscillation, with corresponding eigenvalues $\lambda_j= -\gamma_j/2-i \delta\omega_j$ that are written in terms of the collective resonance frequencies $\delta \omega_j$ (the shift of the resonance frequency with respect to the arc frequency $\omega_0$) and decay rates $\gamma_j$. The changes of $\gamma_j$ represent collective enhancement of radiation when $\gamma_j$ is larger than the decay rate of an isolated meta-atom (superradiance) and collective suppression of radiation in the opposite case (subradiance)~\cite{Dicke54}.

Although in the experiments it is not practical to ensemble-average over a large number of realizations of disorder, in numerical simulations we can fully analyze the statistical properties of the electromagnetic (EM) response due to disorder in the positions of the scatterers.
For each individual stochastic realization of meta-molecule positions, we calculate the EM response for the quantities of interest. By means of ensemble-averaging over many such realizations, we obtain both the averages and statistical fluctuations of the EM response of the magneto-dielectric array \cite{Jenkins2018}. For  displacement $\delta\spvec{r}_\ell$  of the unit cell and an observable quantity $O$ of an array of $N$ ASR resonators we then obtain
  \begin{align}
    \label{eq:StochEq1}
    \left\langle O \right\rangle  & = \int d^3\delta r_1\ldots d^3\delta
    r_N\, O(\spvec{r}_1,\ldots,\spvec{r}_N)
    P(\delta\spvec{r}_1,\ldots,\delta\spvec{r}_N)\nonumber\\
    &= \frac{1}{\mathcal{N}}
  \sum_{n=1}^{\mathcal{N}}
  O(\spvec{r}_1^{(n)},\ldots,\spvec{r}_N^{(n)}) \,\textrm{.}
  \end{align}
Here we have taken the displacements to be independent and random for each unit cell that simplifies the joint
probability distribution $P$ for displacements $\delta\spvec{r}_\ell$ of ASRs, and we also assume the displacements to be uniformly distributed within the square interval $x \in (-aD/2, aD/2)$, $y \in (-aD/2, aD/2)$, where $a$ is the periodic array unit cell size and $D$ quantifies the strength of disorder. We similarly calculate the statistical variances
 \begin{equation}
  \label{eq:mc_vari}
(\Delta O)^2= \left< O^2 \right> - \left<O\right>^2  \,.
\end{equation}

\section{Experimental Methods}
\subsection{Samples}
Periodic and disordered ASR metamaterials were fabricated by etching a 35 $\mu$m copper film on a 1.6 mm thick dielectric (FR4) substrate. The inner (outer) radius of the ASR resonators was 2.8 mm (3.2 mm). The arrays comprised a grid of $30\times 36$, where the lattice spacing was $a = 7.5$ mm in the periodic sample. In the disordered samples, the center of each meta-molecule was displaced following a random uniform distribution defined over the square interval $x \in (-aD/2, aD/2)$, $y \in (-aD/2, aD/2)$, where $D=0.22$ is the degree of disorder.

\subsection{Near-field measurements}
The near-field response of the metamaterial arrays was characterized by a microwave near-field scanning microscope embedded in an anechoic chamber \cite{SavoEtAlPRB2012}. The samples were illuminated by a horn antenna with the electric field oriented along the arcs of the ASRs (parallel to the y-axis of Fig.~\ref{fig:schematic}b). A $2.5$ mm electric monopole antenna collected the electric field component normal to the array plane at a distance of  $\sim 1$ mm from the array and the signal was recorded by the vector network analyser. For each sample, a central area of $20 \times 20$ unit cells was scanned with a step of $0.25$ mm.

\section{Resonator excitations and near fields}
\begin{figure*}
  \centering
  \includegraphics[width=\textwidth]{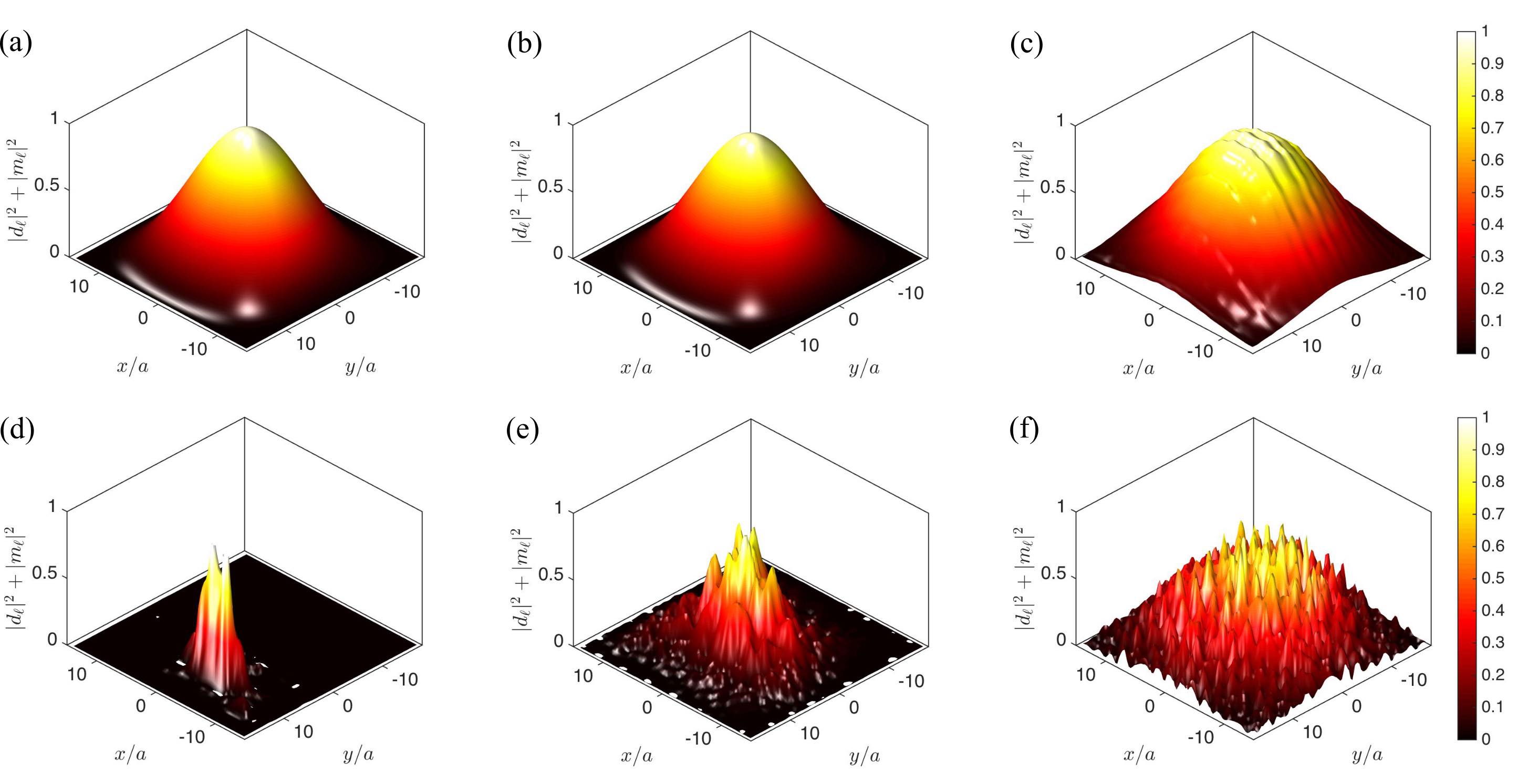}
  \caption{\textbf{The effects of density on the collective uniform magnetic eigenmode.} Excitations of a single eigenmode with vayring lattice spacing $a$ for regular (a-c) and disordered (d-f) ASR arrays. The corresponding lattice spacings are $a=0.28\lambda$ (a,d), $0.83\lambda$ (b,e), and $1.4\lambda$ (c,f), and the radiative decay rates are $0.205\Gamma$ (a), $0.073\Gamma$ (b), $0.909\Gamma$ (c), $0.210\Gamma$ (d), $0.101\Gamma$ (e), and $0.960\Gamma$ (f). In the disordered array, the degree of disorder is $D=0.11$.}
  \label{fig:eigenmodes}
\end{figure*}

The well-known effect of positional disorder in localizing near-field excitations can be linked to the dramatic behavior of the collective eigenmodes of the metamaterial. In the case of regular ASR metamaterial arrays, an incident field can lead to excitations, where the dominant contribution comes from a  single subradiant magnetic eigenmode (see Fig.~\ref{fig:eigenmodes}a) with a suppressed subradiant collective radiative decay rate of $0.205\Gamma$ that extends across the metamaterial array~\cite{Jenkins2017PRL}. Introduction of even moderate disorder leads to a dramatic deformation of the eigenmode, from a spatially extended uniform mode to a strongly localized one (see Fig.~\ref{fig:eigenmodes}d). Such dramatic effects are related to the interplay between the strong collective interactions across the metamaterial array and positional disorder. Indeed, the (dipole-dipole) interactions between the metamaterial resonators depend strongly on the lattice spacing and become weaker as the latter increases, or equivalently, the density of resonators decreases. For instance, when the lattice spacing becomes larger than the wavelength, the collective mode of the periodic array loses its subradiant character and its decay rate almost reaches that of the single resonator decay rate ($\Gamma$) (Fig.~\ref{fig:eigenmodes}c). As a result, disorder now does not lead to a localized subradiant mode as in the case of dense arrays (Fig.~\ref{fig:eigenmodes}d), but rather to a strongly radiating mode with multiple regions of excitation across the array (Fig.~\ref{fig:eigenmodes}f).

\begin{figure*}
  \centering
    \centering
    \includegraphics[width=\textwidth]{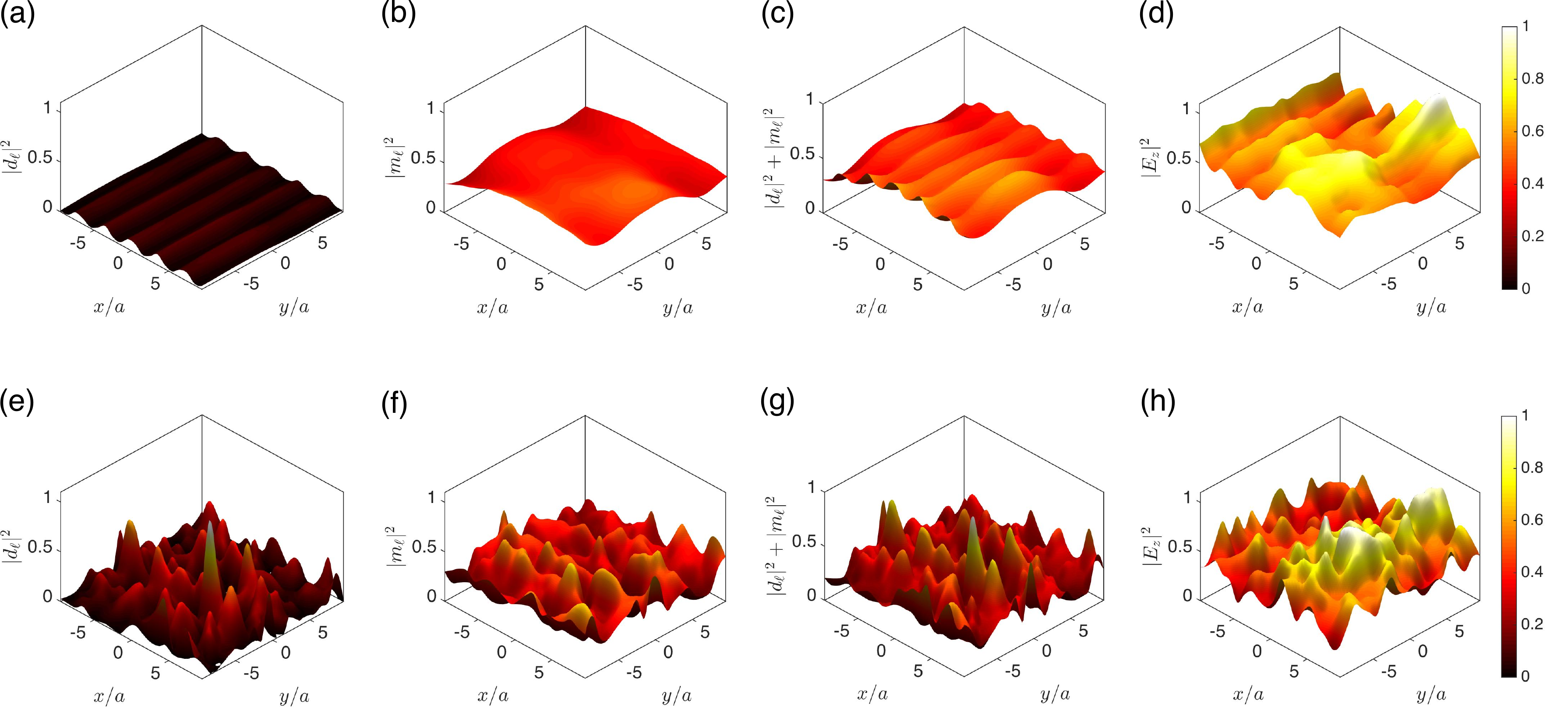}
  \caption{\textbf{Disorder-induced localization in microwave ASR
    arrays.}. Theoretical (a-c \& e-g) and experimental (d \& h) near field maps
    for regular (a-d) and disordered (e-h) arrays under illumination
    with a nearly uniform wavefront as obtained from experimental
    measurements. The theoretical electric
    dipole intensity of each meta-molecule is quantified by $|d_\ell|^2$ (a \& e),  whereas the magnetic dipole intensity by $|m_\ell|^2$ (b \& f).
    Both the theoretical framework and experimental observations
    demonstrate localized regions of field enhancement, which we
    show to exhibit strong magnetic dipole moments. The spacing in the ordered arrays is $a=0.278\lambda$. In the case of disordered arrays, the degree of disorder is $D=0.22$.
  }
  \label{fig:microwave_exp}
\end{figure*}

The effects of interactions and disorder are most prominent in the microwave regime where
we find a qualitative agreement between the theory and experiment. Figure \ref{fig:microwave_exp} shows the distribution of excitations in an array driven on resonance with the uniform magnetic mode of the regular array. For a regular array, the theory predicts
(Figs.~\ref{fig:microwave_exp}a-c) a response that qualitatively agrees with that observed experimentally (Fig.~\ref{fig:microwave_exp}d).
Our model indicates that the disordered metamaterial of
Figs.~\ref{fig:microwave_exp}e-g supports regions in which the
meta-molecular excitations are enhanced by about $80\%$ with respect
to those of a regular array, and shows the same localized pattern of
excitation observed in the experiment (Fig.~\ref{fig:microwave_exp}h).

\begin{figure*}
  \centering
 \includegraphics[width=1.9\columnwidth]{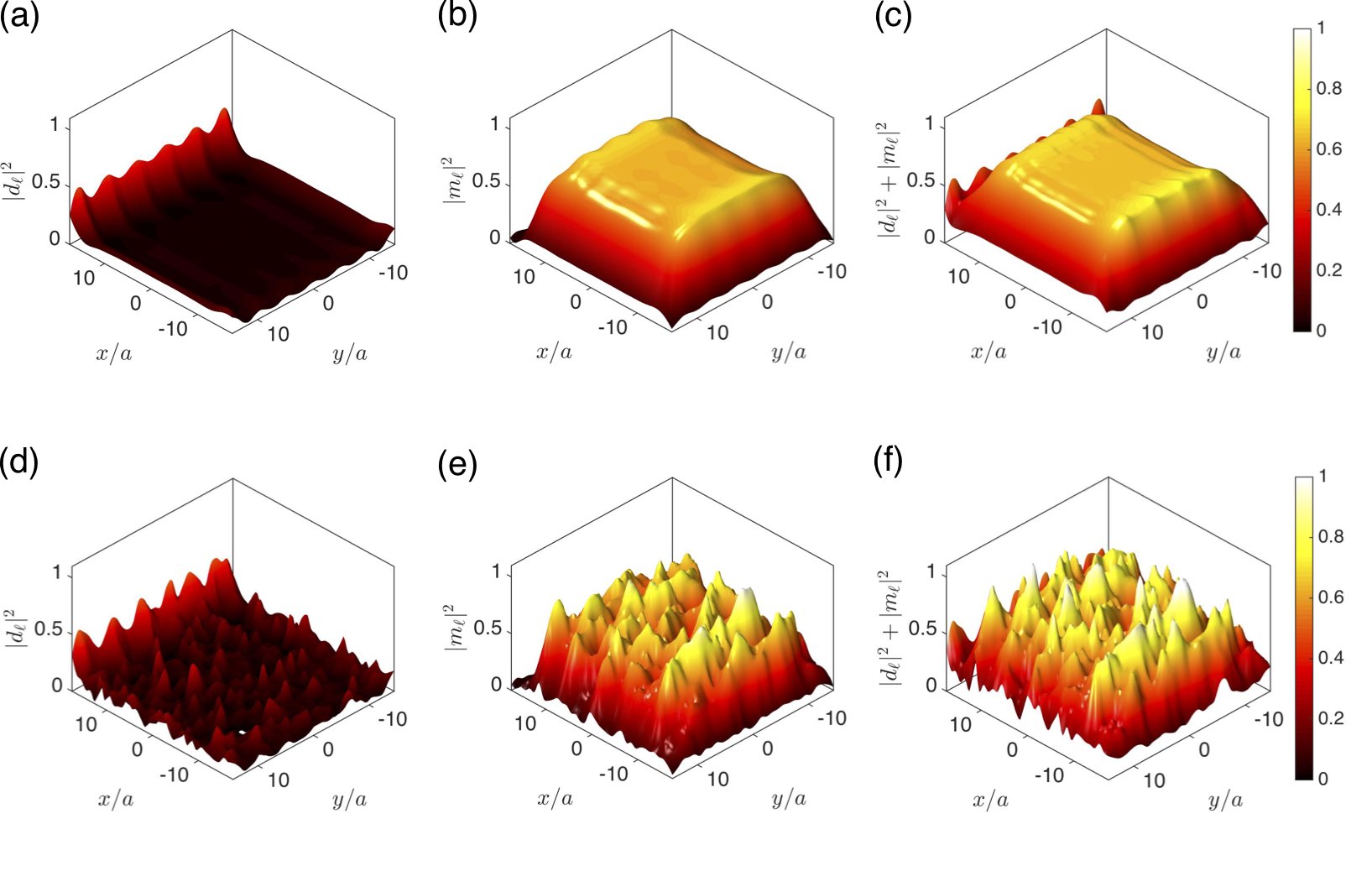}
  \caption{\textbf{Disorder-induced localization in plasmonic metamaterial arrays.} Panels (a-c) show the electric dipole (a), magnetic dipole (b) and total (c) excitations of plasmonic ordered arrays. The corresponding excitations of plasmonic disordered arrays are presented in (d-f). The degree of disorder is $D=0.22$ and the lattice constant $a=0.2\lambda$.}
  \label{fig:optical}
\end{figure*}

In the optical part of the spectrum, metallic metamaterials suffer from Ohmic losses, which limit the role of interactions in
the response. However, we find evidence also in the near field response of the plasmonic metamaterial that the collective phenomena still manifest themselves.
We show in Fig.~\ref{fig:optical} how interactions between plasmonic resonators result in localized regions of the array being more
excited in response to an incident field than any meta-molecule would
be in a regular array.
For a specific configuration of meta-molecule positions ($D=0.22$),
Fig.~\ref{fig:optical}(d-f) shows that certain meta-molecules in the
disordered array (Fig.~\ref{fig:optical}(e)) have magnetic dipole
intensities enhanced by $50\%$
over the most excited magnetic dipole of the regular array
(Fig.~\ref{fig:optical}(b)).
We find that the peak energy in disordered plasmonic arrays (Fig.~\ref{fig:optical}(f)) increases even further owing to an increase in the electric dipole excitations.

\section{Localized response}

\begin{figure}
  \centering
  \includegraphics[width=\columnwidth]{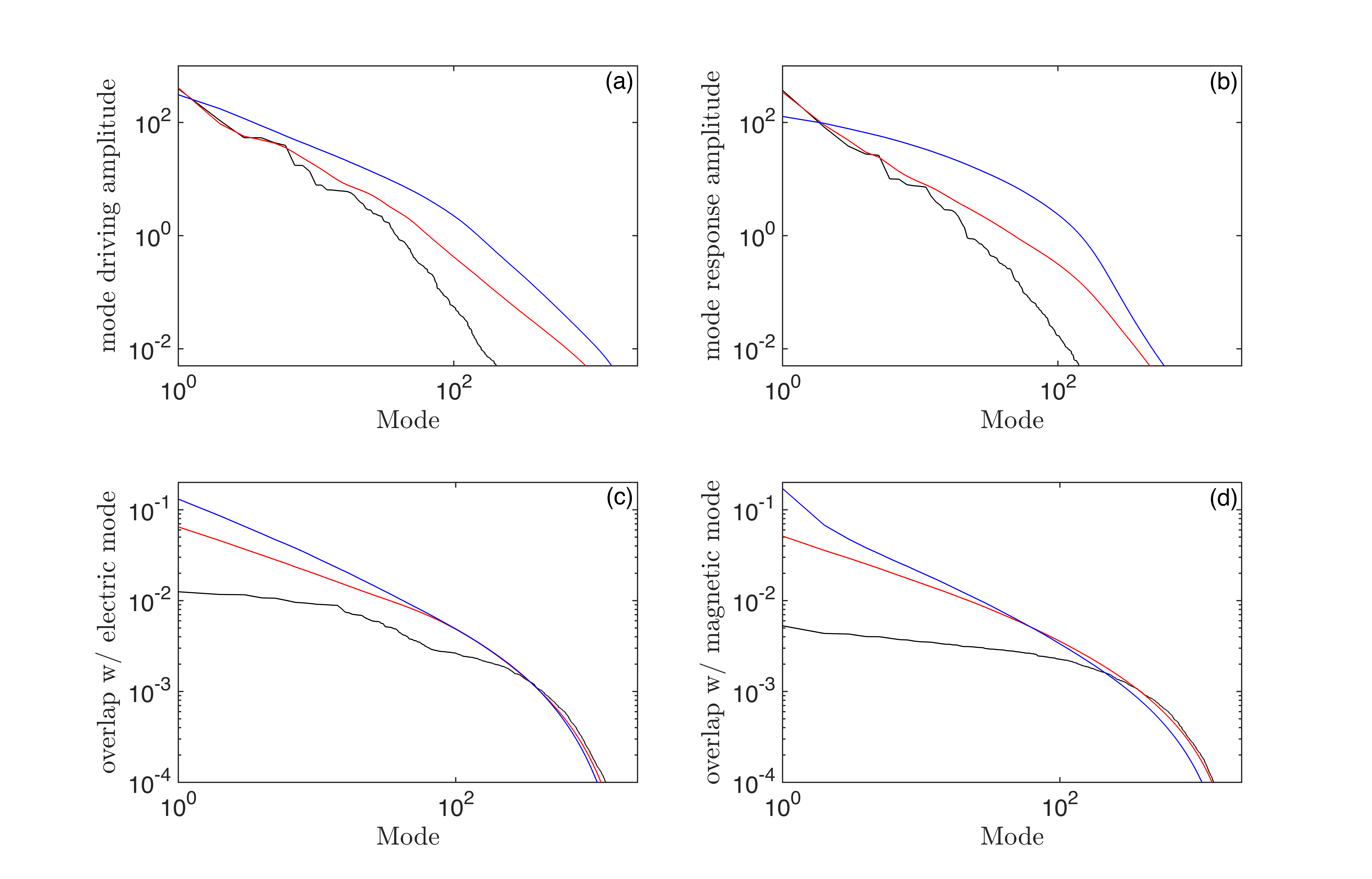}
  \caption{\textbf{Mode characteristics of plasmonic ASR arrays:}
    (a) The incident field excitation (squared) applied to each of the
    collective eigenmodes by a plane wave; (b) average excitation
    intensity of each of the collective modes when the plane wave is tuned to the transmission resonance of a regular array. (c-d) contribution of the collective eigenmodes to an electric (c) or magnetic (d) dipole excitation localized in a single metamolecule. Each of the quantities is calculated for each mode and each realization of metamolecule positions. For each realization, we order the modes by the relevant quantity in decreasing order and compute the average over all realizations. The calculated quantities correspond to disorder parameters $D=0$ (black), $D=0.22$(red), $D=0.44$ (blue).
  }
  \label{fig:sim_modes_excitation_distributions}
\end{figure}

The dramatic changes in the near-field response of metamaterial arrays upon introducing disorder provide opportunities for the engineering of the optical near-field landscape. Indeed, the near-field response of the metamaterial can be traced to the collective radiative eigenmodes supported by the array and their coupling to the incident wave. Here we study the collective excitations of the disordered metamaterial arrays by employing the array eigenmodes as a basis in which we expand both the driving field (plane wave) (see Section~\ref{section:model}) and the response of the array Figs.~\ref{fig:sim_modes_excitation_distributions}(a-b). In the case of a regular array, a plane wave tuned at the transmission resonance can excite strongly only a handful of eigenmodes (black lines) with both the driving and the response amplitude decreasing rapidly for other modes. In fact, in the limit of an infinitely large array, the plane wave would only couple to the two eigenmodes in which the array oscillates uniformly. On the other hand, with increasing disorder the number of eigenmodes that are excited increases rapidly and, in the case of strong disorder (blue lines), as many as 100 eigenmodes can be strongly excited. As an example of near-field engineering, we consider the preparation of strongly localized excitations in the array. In Figs.~\ref{fig:sim_modes_excitation_distributions}(c-d) we calculate the contributions of different collective eigenmodes in achieving an electric or magnetic dipole excitation, respectively, localized in a single unit cell for varying degree of disorder. In both the electric and magnetic dipole case, achieving a localized excitation in regular arrays requires a large number of collective modes. This is a direct result of the extended character of the eigenmodes in regular arrays. Conversely, in disordered arrays, the collectives eigenmodes become increasingly localized (see Fig.~\ref{fig:eigenmodes}). Hence, the number of modes required decreases substantially, and in the case of strong disorder, a handful of eigenmodes suffices to form localized excitations (see blue lines in Figs.~\ref{fig:sim_modes_excitation_distributions}(c-d)). This localization of collective eigenmodes occurs in the plane of the metamaterial array and is drasticallty different from the localization of waves propagating in 1D media \cite{Nori2008, Gredeskul2012, Oliveira2011}.

\begin{figure}
  \centering
  \includegraphics[width=\columnwidth]{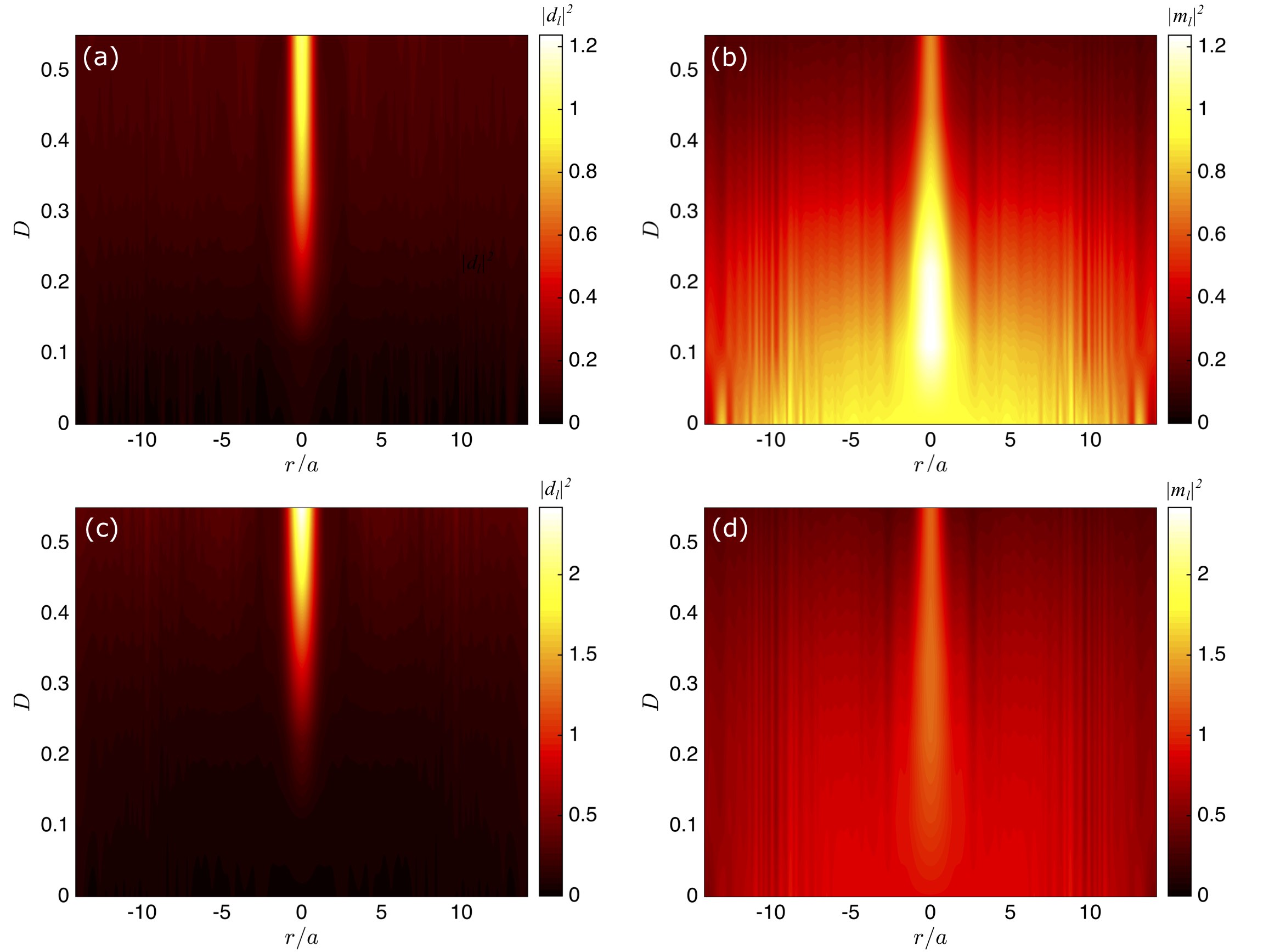}
  \caption{\textbf{Localized meta-molecule excitations.}
  Electric (a,c) and magnetic (b,d) dipole excitations of the most strongly excited meta-molecule and its neighbouring metamolecules for microwave (a-b) and plasmonic (c-d) arrays as a function of the disorder parameter $D$. The colormaps present averages over all realizations and are symmetric around $r=0$.}
  \label{fig:I_max_corr_plot_optics}
\end{figure}

\begin{figure}
  \centering
  \includegraphics[width=\columnwidth]{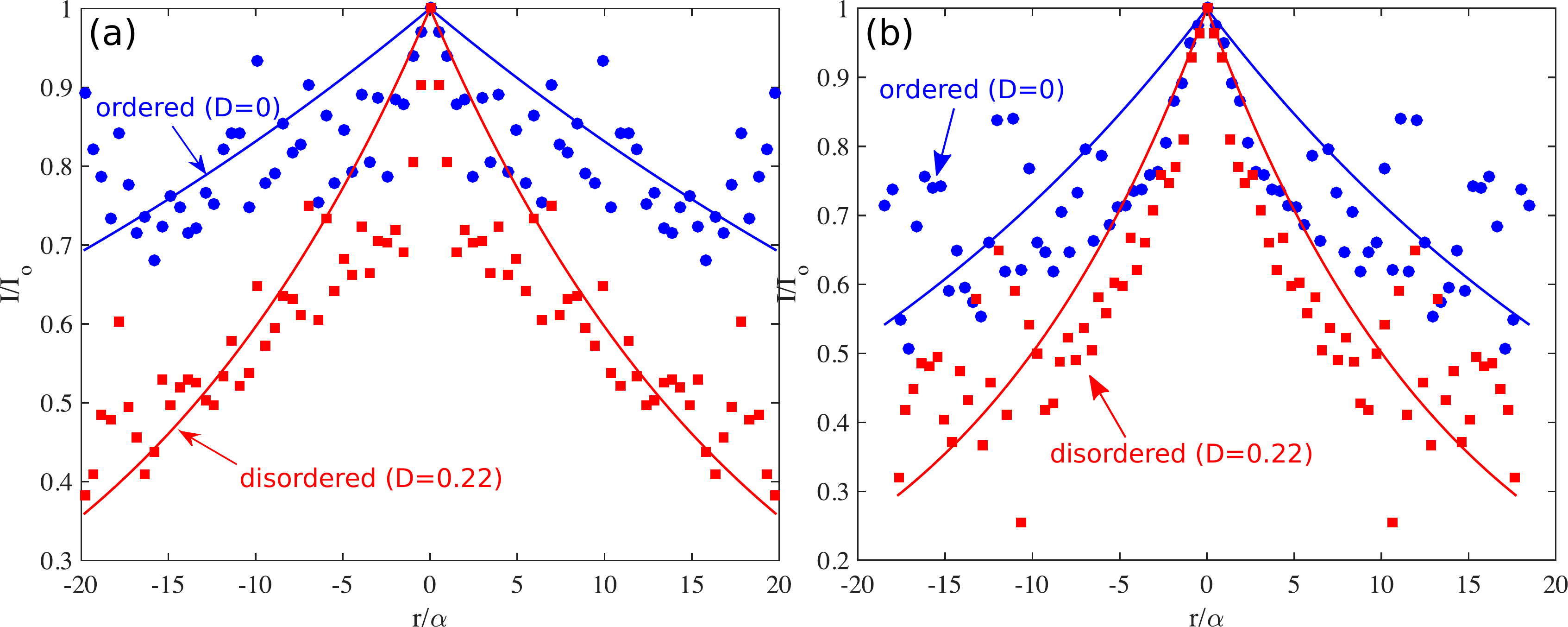}
  \caption{\textbf{Examples of microwave localization.}
Experimentally measured electric field intensity profiles around characteristic positions in disordered ($D=0.22$) and ordered ASR arrays. Panel (a) shows the field profile of a strongly confined excitation in the disordered array (red) and around the same position in the ordered array (blue). Solid lines correspond to fits of the form $I/I_o=e^{-|r|/r_o}$, where $I_o$ is the electric field intensity at the position of interest, $r_o=54\alpha$ for the periodic array (blue) and $r_o=19\alpha$ for the disordered one (red). Panel (b) presents the field profile around the most strongly excited metamolecule in the disordered (red) and ordered (blue) array with corresponding fitting parameters  $r_o=14\alpha$ and $r_o=30\alpha$, respectively. All graphs are symmetric around $r=0$.
  \label{fig:exp_loc}}
\end{figure}

The large number of eigenmodes that are accessible in disordered metamaterial arrays can be employed to engineer strongly localized excitations. To determine the extent of this localization, we numerically simulate the collective response of $1024$ disordered microwave arrays for varying degrees of disorder. For every realization of meta-molecule positions, we consider the region within ten unit-cells of the most excited meta-molecule (excluding meta-molecules close to the array edges). Following an averaging process over all realizations, the electric and magnetic dipole excitation of this meta-molecule and its vicinity are presented in Figs.~\ref{fig:I_max_corr_plot_optics}(a-b), respectively. In all cases, regular arrays exhibit an absence of localized excitations with slow variations of intensity across the array. However, the introduction of disorder leads to increasingly localized excitations that extend over a handful of unit cells. With increasing degree of disorder, the electric dipole intensity increases rapidly and begins to saturate at $D=0.5$, while the spatial extent of the localized excitation remains constant. On the other hand, the magnetic dipole intensity initially increases with disorder up to $D\simeq 0.15$ and then decreases, while its size decreases continuously with disorder. The situation is very similar in the case of plasmonic arrays (Figs.~\ref{fig:I_max_corr_plot_optics}(c-d)) with the magnetic dipole intensity peaking at $D\simeq 0.3$. In both the microwave and plasmonic cases, the localized excitation for ordered and weakly disordered arrays is predominantly of magnetic dipole character, but at strong disorder it quickly converts to electric dipole. This behavior is corroborated by the experimental measurements of disordered ASR metamaterials. A typical example is presented in Fig.~\ref{fig:exp_loc}a, where we plot the electric field intensity in the vicinity of a localized excitation for a disordered ASR array with $D=0.22$ (red squares). The excitation is primarily localized in a small number of unit cells, while it decays rapidly away from its centre. In comparison, the same area in a regular array (blue circles) exhibits an almost flat profile. Similar behaviour is observed around the most strongly excited metamolecule in the arrays (see Fig.~\ref{fig:exp_loc}b), with the excitation in the disordered array being substantially more confined than in the case of the ordered array. In fact, the variation in the field distribution of the ordered array is attributed mainly to the inhomogeneity of the incident wave.

The ability to generate localized excitations in plasmonic metamaterial arrays can be exploited to strengthen the coupling between material excitations and, e.g., quantum emitters in order to control the decay rate of the latter. In fact, each collective eigenmode of the metamaterial array can act as an effective cavity whose quality factor is linked to the collective eigenmode decay rate. Thus, collective meta-molecule excitations can serve as an intermediary for an external field to strongly drive quantum emitters. In contrast to the localization observed in random metal/dielectric composites at the percolation threshold consisting of non-resonant inclusions \cite{Shalaev1999, Shalaev2001}, the advantage of the ASR metamaterial is that the enhancement can be achieved with a prescribed multipole (magnetic or electric dipole) character.

\begin{figure}
  \centering
  \includegraphics[width=\columnwidth]{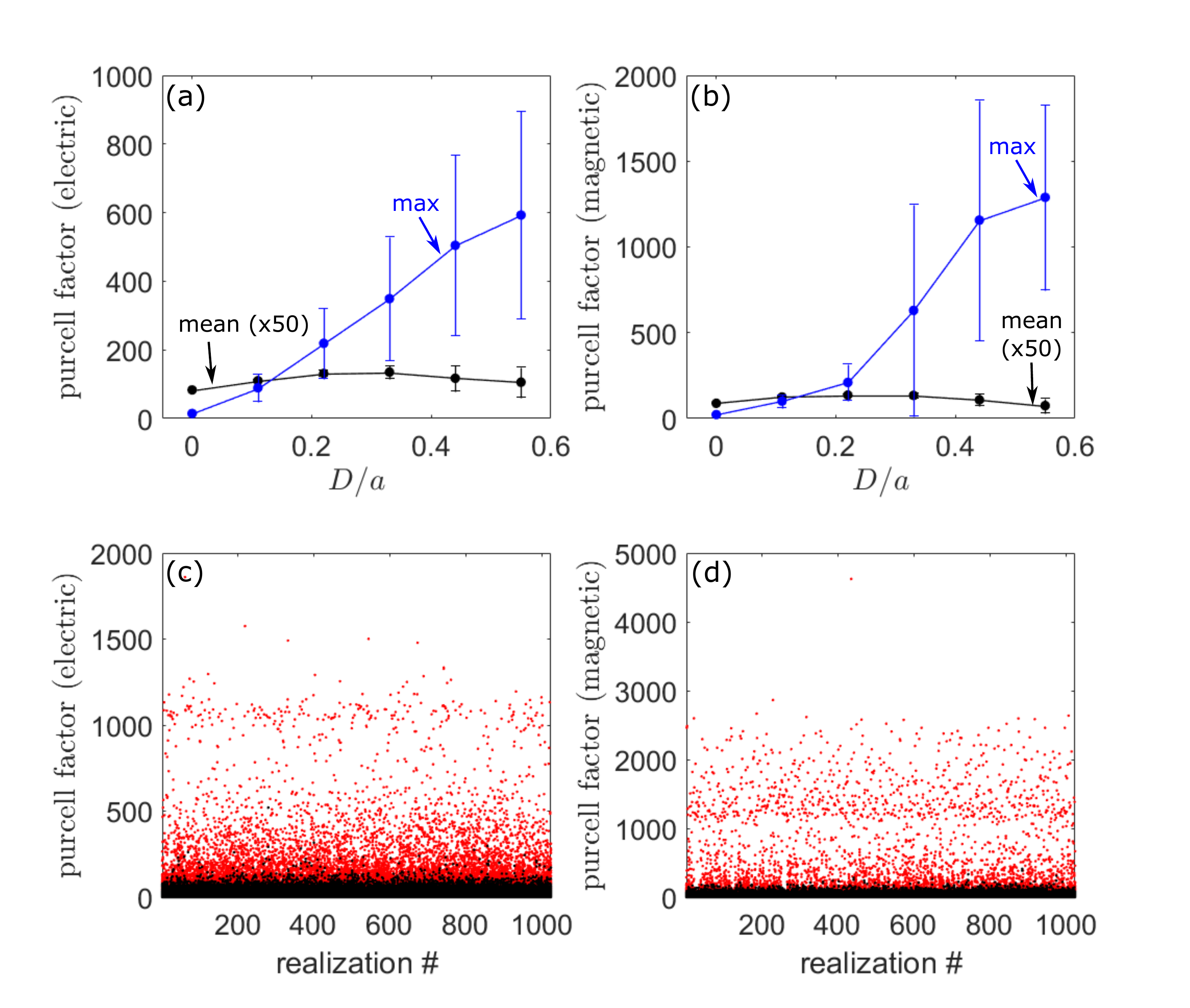}
  \caption{\textbf{Purcell enhancement in plasmonic metamaterial arrays.} (a-b) Average (black) and maximum (blue) Purcell factor for electric (a) and magnetic (b) dipole excitations as a function of disorder. The average is calculated over 1024 different realizations and over all modes of each realization, and it has been multiplied by a factor of 50. The maximum Purcell factors have been calculated by finding the maximum value for each realization and then averaging over all 1024 realizations. The errorbars correspond to the standard deviation of the average and maximum Purcell factor across different realizations. (c-d) Purcell factors for electric (c) and magnetic (d) dipole excitation for different realizations and for two different degrees of disorder: 0.11 (black) and 0.55 (red). Each point in the graphs corresponds to the Purcell factor of a single mode of a single realization.
  }
  \label{fig:purcell}
\end{figure}

Here we calculate the Purcel factors and their statistical distributions for the collective modes of the array normalized to that of a single arc (see App.~\ref{appendix:purcell}). We demonstrate that coupling to a single collective mode can enhance the emitter's decay rate by more than three orders of magnitude. In Figs.~\ref{fig:purcell}a\&b, we present the Purcell factor as a function of the degree of disorder for eletric dipole and magnetic dipole excitations, respectively. Whereas the average over all modes and realizations depends weakly on disorder, the maxima (averaged over all realizations) of the Purcell factor increase monotonically with increasing degree of disorder for both electric and magnetic dipole excitations. At the same time, the standard deviation of the maxima Purcell factor values (represented by the error bars in Figs.~\ref{fig:purcell}a\&b) also increases substantially. This indicates that as disorder increases it becomes increasingly likely that there is at least one collective eigenmode that can substantially enhance the Purcell factor. This behavior is further illustrated in Figs.~\ref{fig:purcell}c\&d, where the Purcell factor for each collective eigenmode and each realization is presented for electric and magnetic dipole excitations, respectively. In the case of weakly disordered metamaterial arrays (black points), the Purcell factor values for all modes and realizations are similar both for electric and magnetic dipole excitations. However, as disorder increases (red points), the distribution of Purcell factors of each realization becomes much broader with very high values becoming increasingly more likely.

\section{Concluding remarks}
Strong EM field-mediated interactions between metamolecules can lead to a collective response, where simple homogeneous medium descriptions no longer are valid~\cite{Jenkins2017PRL}. As the electrodynamic behaviour of the metamaterial is then determined by collective excitation eigenmodes,the near-field response under excitation with a delocalized field can be localized even in the absence of disorder~\cite{Sentenac2008,KAO10}, or similarly delocalized in the presence of a localized field excitation~\cite{AdamoEtAlPRL2012}. Here we have shown
that combining positional disorder and strong field-mediated interactions leads to a more complex interplay between collective eigenmodes and the near-field effects.

Our work, in particular, presents an analysis of the cooperative response of disordered metamaterials that allows to tailor the metamaterial near-field landscape with application in the design of artificial EM materials and devices. Controlled localization in metamaterials holds the potential for random lasing~\cite{WiersmaNatPhys2008} or disorder-enhanced nanoantennas, where the electric or magnetic dipole field can be selectively enhanced allowing thus for engineering of the decay rate of emitters positioned in the vicinity of the metamaterial. Our approach is also suitable for driving high-order multipole emitters, which are typically weak but technologically relevant~\cite{HeinGiessenPRL2013}. For instance, disordered ASR arrays provide enhanced localized magnetic dipole excitations, while similar effects can be achieved for higher order terms of the multipole expansion by similar methods~\cite{Watson2016,Watson2017}. Moreover, engaging collective modes in metamaterial arrays does not only allow control the emitter decay rate and multipole character, but could also enable the control of the wavefront and direction of emitted radiation. Other promising applications include novel platforms for sensing, nonlinear optics, focusing, and even cavity quantum electrodynamics, allowing coherent Rabi oscillations between atomic excitations and collective meta-molecular excitations.

\begin{acknowledgments}
We acknowledge financial support from the EPSRC (EP/G060363/1, EP/M008797/1), the Leverhulme Trust, the Royal Society, and the MOE Singapore
Grant No. MOE2011-T3-1-005. We also acknowledge the use of the IRIDIS
High Performance Computing Facility at the University of
Southampton.
\end{acknowledgments}

\appendix

\setcounter{equation}{0}
\setcounter{figure}{0}
\renewcommand{\theequation}{A\arabic{equation}}
\renewcommand{\thefigure}{A\arabic{figure}}

\section{Purcell factor}
\label{appendix:purcell}
In an ideal case, the Purcell factor is estimated by considering an emitter placed in the vicinity of the metamaterial array which couples to a number of different collective modes of the array. We can then attribute a Purcell factor to each of these modes. Here, we approximate this mode specific Purcell factors by the following procedure. For each realization, we find the most excited metamolecule under plane wave normal incidence illumination. We then assume an either purely electric or purely magnetic dipole excitation localized in this metamolecule. This localized excitation can be expanded to the eigenmodes of the system with amplitudes $u_n^{(e)}$ and $u_n^{(m)}$, where $n$ is a mode index and $e,m$ refer to electric or magnetic dipole excitations. Since the Purcell factor depends on the effective volume (or surface in the case of planar metamaterial arrays) of the mode, here we estimate the number of unit cells across which a mode is spread as $|u_n^{(e)}|^{-2}$ and $|u_n^{(m)}|^{-2}$. Assuming an effective cavity with reference surface $A_0$ and decay rate $\Gamma$ corresponding to a single arc, we can write the Purcell factor for eigenmode $n$ in the rotating wave approximation as:
\begin{equation}
  P_n^{(e/m)}=P_0\frac{|u_n^{(e/m)}|^2}{\gamma_n/\Gamma}
  \label{eq:purcell}
\end{equation}
where $P_0=6\pi c^3/(A_0^2\Omega_0^2\Gamma)$ and $\gamma_n$ is the decay rate of mode $n$.


%

\end{document}